\documentclass[10pt,twocolumn,aps,prl,showkeys, showpacs,groupedaddress]{revtex4}
\usepackage{epsfig}

\begin{document}

\title{Thermoelectric power of Ba(Fe$_{1-x}$Co$_x$)$_2$As$_2$, $0\le x\le 0.05$, and Ba(Fe$_{1-x}$Rh$_x$)$_2$As$_2$, ($0\le x \le 0.171$)}.

\author{H. M. Hodovanets, A. Thaler, E. Mun\footnote{Currently at the  National  High Magnetic Field Laboratory at Los Alamos National Laboratory}, N. Ni\footnote{Currently: Marie Curie Fellow at Los Alamos National Laboratory}, S. L. Bud'ko and P. C. Canfield}

\affiliation{Ames Laboratory and Department of Physics \& Astronomy, Iowa State University, Ames, IA 50011, USA}

\begin{abstract}
Temperature-dependent, in-plane, thermoelectric power data are presented for single crystals of Ba(Fe$_{1-x}$Co$_x$)$_2$As$_2$, ($0\le x \le 0.05$), and  Ba(Fe$_{1-x}$Rh$_x$)$_2$As$_2$, ($0\le x \le 0.171$).  Given that previous thermoelectric power and angle resolved photoemission spectroscopy studies of Ba(Fe$_{1-x}$Co$_x$)$_2$As$_2$ delineated a rather large Co-concentration range for Lifshitz transitions to occur, and the underdoped side of the phase diagram is poorly explored, new measurements of thermoelectric power on tightly spaced concentrations of Co, $0\le x \le 0.05$, were carried out. The data suggest evidence of a Lifshitz transition, but instead of a discontinuous jump in thermoelectric power between $0\le x \le 0.05$, a more gradual evolution in the S(T) plots as $\it x$ is increased was observed. The thermoelectric power data of Ba(Fe$_{1-x}$Rh$_x$)$_2$As$_2$ show very similar behavior to that of Co substituted BaFe$_2$As$_2$. The previously outlined $\it T - x$ phase diagrams for both systems are further confirmed by these thermoelectric power data.
\end{abstract}


\maketitle

There are models of superconductivity (SC) for Fe-As superconductors in which details of the band structure/ Fermi surface are considered important \cite{Mazin,Kemper,Hirschfeld,Chubukov}. For the Ba(Fe$_{1-x}$Co$_x$)$_2$As$_2$ family, angle-resolved photoemission spectroscopy (ARPES) measurements observed that the onset and the offset of superconductivity occurs in proximity to electronic topological transitions, $\it i. e.$, Lifshitz transitions  \cite{Liu, Liu1}, suggesting that SC is connected to the changes in the Fermi surface topology. In this context, then, it is worth recalling that thermoelectric power (TEP) is very sensitive to the changes in the Fermi surface topology \cite{Varlamov,Blanter,Lifshitz}. Recently we have presented TEP data on Ba(Fe$_{1-x}$Co$_x$)$_2$As$_2$  \cite{Eundeok1} that showed a sudden change in  TEP between $\it x$=0.02 and $\it x$=0.024 over the whole measured temperature range. This feature was atributed to a Lifshitz transition, this conclusion was  supported by subsequent ARPES measurements \cite{Liu}.  In order to better understand the nature and details of this feature, we carried out TEP measurements on tightly spaced Co substitution levels so as to look in more detail at the region of Co substitution where the Lifshitz transition appears to occur. 

In addition to a detailed study of TEP in the Ba(Fe$_{1-x}$Co$_x$)$_2$As$_2$ series, we present data on the  Ba(Fe$_{1-x}$Rh$_x$)$_2$As$_2$ series as well. Co and Rh belong to the same column of the Periodic Table and it has been shown \cite{Ni1}, that the temperature dependent resistance and magnetization data for Ba(Fe$_{1-x}$Rh$_x$)$_2$As$_2$ are strikingly similar to those of Ba(Fe$_{1-x}$Co$_x$)$_2$As$_2$ resulting in $\it T-x$ phase diagrams that are virtually identical.  In this work we measured TEP of the Ba(Fe$_{1-x}$Rh$_x$)$_2$As$_2$ series, in order to compare the effects of Rh substitution to those of Co on the temperature-dependent TEP and position/manifestation of the Lifshitz transition.

\section{Experimental Methods}

Single crystals of Ba(Fe$_{1-x}$Co$_x$)$_2$As$_2$, ($0\le x \le 0.05$), and Ba(Fe$_{1-x}$Rh$_x$)$_2$As$_2$, ($0\le x \le 0.171$) were grown out of self flux using conventional high temperature solution growth technique described elsewhere \cite{Ni1,Ni,Canfield}. The actual Co and Rh concentrations in the single crystals were determined by using wavelength dispersive x-ray spectroscopy (WDS) in the electron probe microanalyzer of a JEOL JXA-8200 electron-microprobe. All Co and Rh concentrations used in the text are the experimentally measured ones.

The samples were characterized by resistance and magnetization measurements.
A standard 4-probe geometry, $\it ac$ technique was used to measure the electrical resistance of the samples. Electrical contact to the samples was made with platinum wires attached to the samples using EpoTek H20E silver epoxy. The measurements were performed in a Quantum Design Physical Property Measurement System PPMS-14. Magnetic measurements were carried out in a Quantum Design Magnetic Property Measurement System (MPMS) SQUID magnetometer with a magnetic field of 5.5 T applied perpendicular to the c-axis.

The thermoelectric power (TEP) measurements were performed by a {\it dc}, alternating heating (two heaters and two thermometers) technique \cite{Eundeok} using a Quantum Design PPMS to provide the temperature environment between 2 K and 300 K. The samples were mounted directly on the gold plated surface of the SD package of the Cernox thermometers \cite{LakeShore} using  DuPont 4929N silver paste to ensure thermal and electrical contact both of which are very important for the TEP measurements. 

\begin{figure}[tbh]
\centering
\includegraphics[width=1\linewidth]{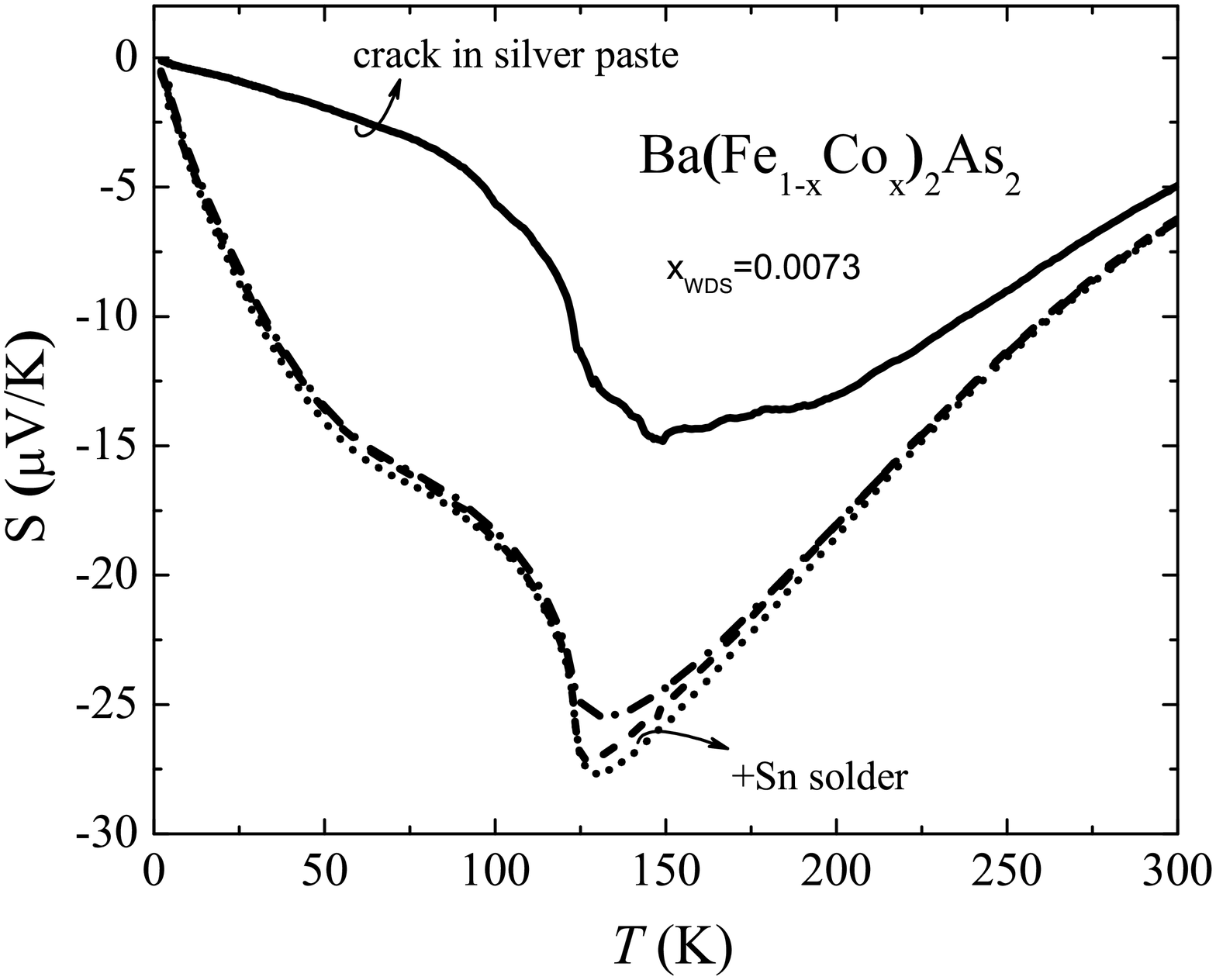}
\caption{\footnotesize In-plane TEP data of Ba(Fe$_{1-x}$Co$_x$)$_2$As$_2$, $\it x$=0.0073. All samples are from the same batch. "+Sn" (short dot curve) - the ends of the sample where dipped into Sn solder to ensure better thermal and electrical contact and then the sample was mounted on the surface of SD package of the Cernox thermometers as explained in the text. Effect (if any) of the Sn on the TEP signal was not corrected for.}
\label{fig:x0.0073}
\end{figure}

While performing these measurements, we found that samples sometimes developed cracks in the silver paste which we believe led to a decreased thermal contact between the sample and the surface of the SD package of the Cernox thermometer and, as a result, somewhat noisier data with a consequently smaller absolute value of the TEP over the whole temperature range: since the heater is mounted on the surface of the SD package of the Cernox thermometer, the actual $\Delta T$ across the sample was probably smaller than that read by the sensors leading to a smaller thermoelectric voltage $\Delta V$ generated. (Given poor contact on one or both sides of the sample, $\Delta V$ should not change dramatically due to zero electrical current but $\Delta T$ should change due to finite heat flow.) 

The TEP data for the sample with the crack in the silver paste and a few other from the same batch for which the silver paste contacts showed no evidence of degradation, are shown in Fig. 1. The data for the sample with the crack in the silver paste is shown in solid black curve, it is clearly noisier data with the absolute value of the TEP significantly smaller compared to the absolute value of the TEP of the other three samples with good thermal contact. For one of the samples, data set presented in small dots, the ends of the sample where dipped into Sn solder to ensure better thermal and electrical contact and then the sample was mounted on the surface of the SD package of the Cernox thermometer as explained above. Additional contributions (if any) of the Sn to the TEP signal were not corrected for. Unfortunately, a simple scaling of the data set for the sample with the crack in the silver paste, solid black curve, to the other data sets does not work most likely because the temperature dependence of the thermal conductivity associated with the degraded contact is not measured or modeled.

Based on these observations, in the cases when multiple TEP measurements on  samples from the same batch had data spread significantly beyond the expected experimental errors, within a maximum of 10$\%$ \cite{Eundeok}, the data set with the largest absolute value of the TEP and highest signal-to-noise ratio was taken as a more reliable measurement. The data for the sample with the crack in the silver paste are regrettably similar to the curves for $\it x$=0.013 and $\it x$=0.02 in our previously published data \cite{Eundeok1}. This realization also motivated us to revisit the data for the low Co-concentration. 

\begin{figure}[tbh]
\centering
\includegraphics[width=1\linewidth]{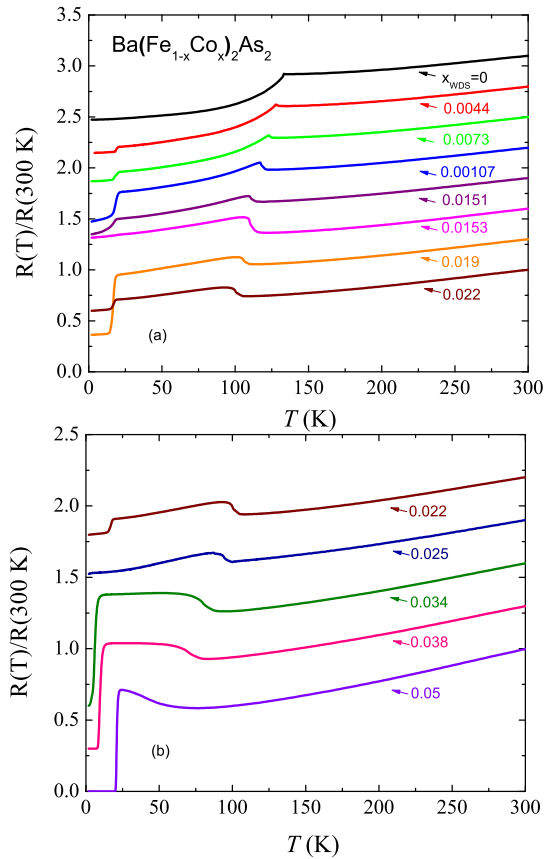}
\caption{\footnotesize (Color online) (a) and (b) Temperature dependence of normalized resistance, R($\it T$)/R(300K), of Ba(Fe$_{1-x}$Co$_x$)$_2$As$_2$, ($0\le x \le 0.05$), single crystals. $\it x$=0.022 is shown in both figures for clarity.  Subsequent data sets are shifted upward by 0.3 for clarity.}
\label{fig:Resistance}
\end{figure}

\begin{figure}[tbh]
\centering
\includegraphics[width=1\linewidth]{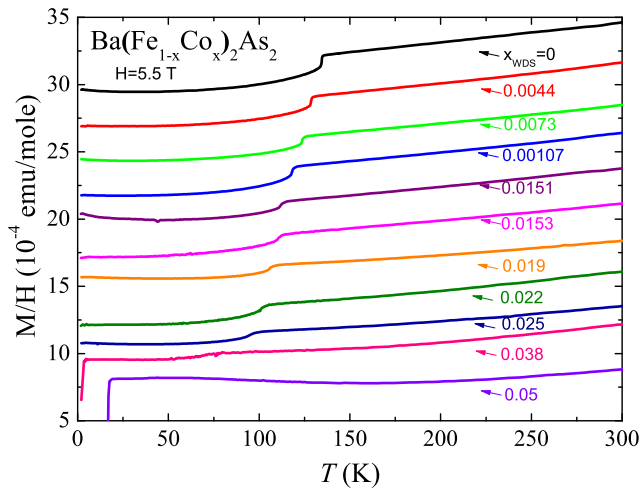}
\caption{\footnotesize (Color online)  Temperature dependence of magnetization of Ba(Fe$_{1-x}$Co$_x$)$_2$As$_2$, ($0\le x \le 0.05$), single crystals. M($\it T$)/H data taken at 5.5 T with H $\perp c$. Subsequent data sets are shifted upward by 2.5$\times 10^{-4}$ emu/mole for clarity.}
\label{fig:Magnetization}
\end{figure}

\section{Data presentation}

The temperature dependent, normalized, resistance data, R($\it T$)/R(300K), and temperature dependent magnetization data for Ba(Fe$_{1-x}$Co$_x$)$_2$As$_2$, ($0\le x \le 0.05$), samples are shown in Figs. 2 and 3 respectively. As Co concentration is increased, the structural, $ T_s$, and magnetic, $ T_m$, transitions \cite{Ni,Canfield} are suppressed and separated, and the sharp feature associated with structural and magnetic transitions broadens. For some of the R($\it T$) data a  kink like feature arises at low temperature resembling a superconducting transition. This kink like feature resembles the pressure-induced, granular or filamentary, superconducting behavior seen in SrFe$_2$As$_2$ \cite{Colombier,Saha} and CaFe$_2$As$_2$ \cite{Torikachvili}, $\it i. e.$, superconductivity is present in a small fraction of the sample, most likely due to internal strains introduced by cleaving and shaping of the samples. The feature associated with the filamentary SC was not observed in high field (5.5 T) magnetization data for the samples from the same batches for which the kink like feature in resistance data was present. In addition, zero field cooled low field (50 Oe) magnetization (not shown here) for 0.0044$\le x \le$0.0107 did not have a feature associated with filamentary SC either. Although the kink like feature  due to filamentary superconductivity was observed at low temperature in the R($\it T$)/R(300K), Fig. 1 shows no SC (no jump to zero TEP) in temperature dependent TEP for $\it x$=0.0073 of Co substitution level. As a matter of fact, none of the Ba(Fe$_{1-x}$Co$_x$)$_2$As$_2$, ($0\le x < 0.034$), samples studied by TEP showed SC or any feature associated with filamentary SC seen in resistance measurements (see below).

The temperature dependent normalized resistance and magnetization measurements for the samples/batches of the Ba(Fe$_{1-x}$Rh$_x$)$_2$As$_2$, ($0\le x \le 0.171$), used in this work, were published earlier \cite{Ni1}. As Rh concentration is increased the structural and magnetic transitions are suppressed to the lower temperatures in a manner similar to Co substitution. 

\begin{figure}[tbh]
\centering
\includegraphics[width=1\linewidth]{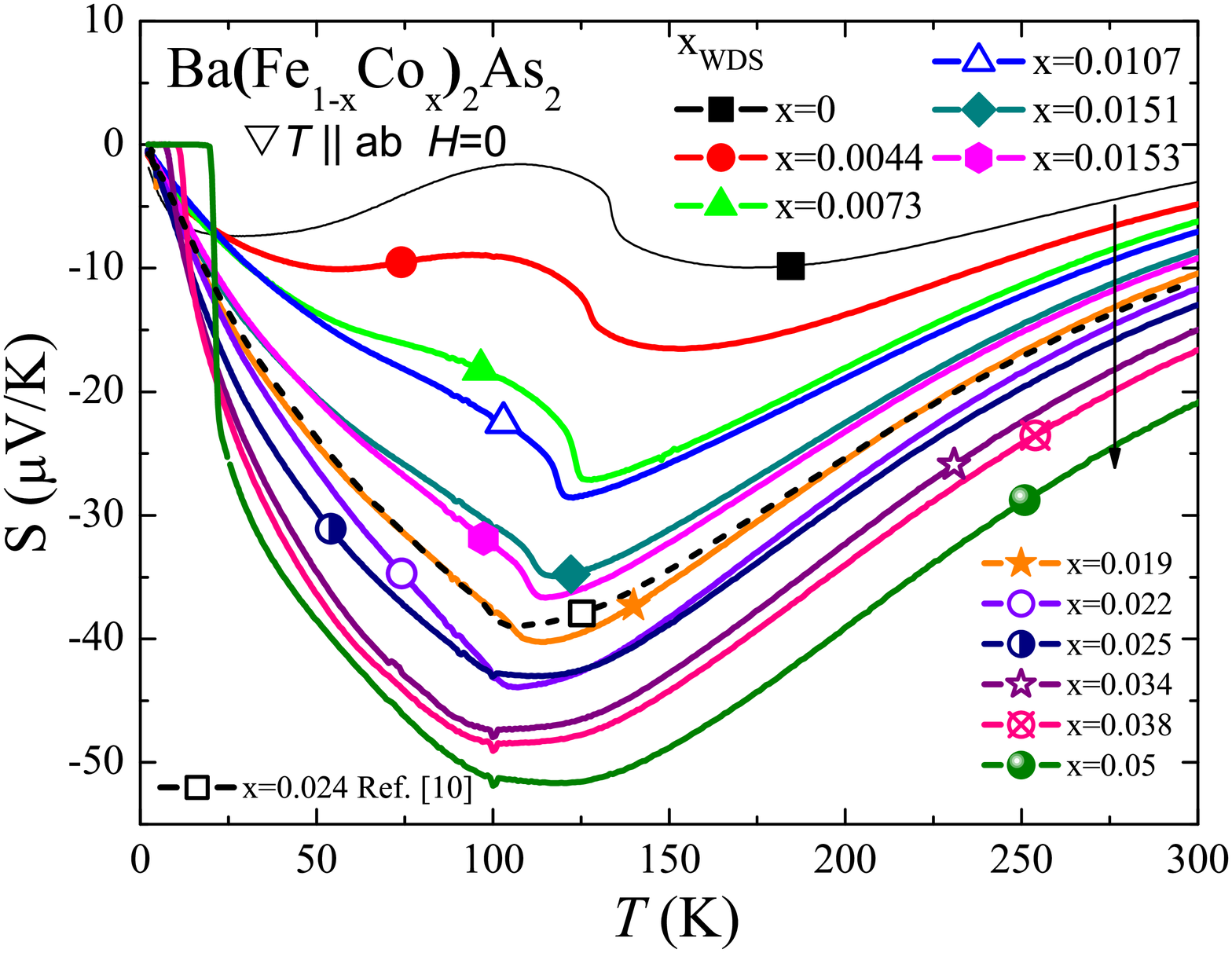}
\caption{\footnotesize (Color online) In-plane temperature dependent TEP, $\it S$, of Ba(Fe$_{1-x}$Co$_x$)$_2$As$_2$, ($0\le x \le 0.05$). $\it x$=0.024 and $\it x$=0 are adopted from \cite{Eundeok1}.}
\label{fig:TEP}
\end{figure}

\begin{figure}[tbh]
\centering
\includegraphics[width=1\linewidth]{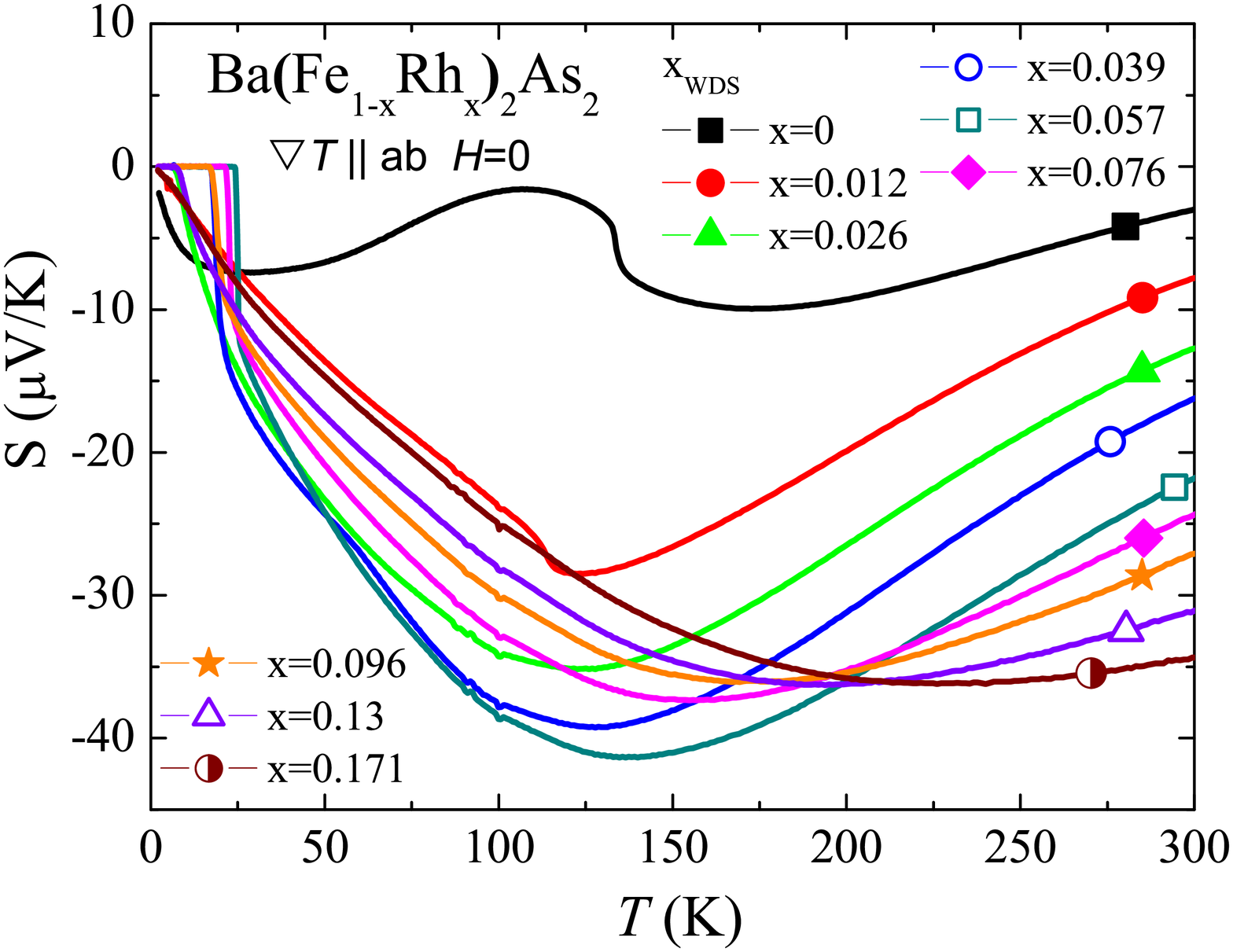}
\caption{\footnotesize (Color online) In-plane temperature dependent TEP, $\it S$ of Ba(Fe$_{1-x}$Rh$_x$)$_2$As$_2$, ($0\le x \le 0.171$). $\it x$=0 is adopted from \cite{Eundeok1}.}
\label{fig:Rh TEP}
\end{figure}

The temperature dependent TEP data for the newly grown, tightly spaced Ba(Fe$_{1-x}$Co$_x$)$_2$As$_2$ samples, ($0\le x \le 0.05$), are shown in Fig. 4. The data evolve gradually instead of manifesting sudden jump as reported in Ref. \cite{Eundeok1}. The absolute value of TEP increases as Co concentration is increased. The data for Ba(Fe$_{1-x}$Co$_x$)$_2$As$_2$, $\it x$=0 and $\it x$=0.024 are taken from Ref. \cite{Eundeok1} for comparison. SC was not observed for any of the Ba(Fe$_{1-x}$Co$_x$)$_2$As$_2$,  ($0\le x < 0.034$), samples studied by TEP in this work. SC appears for $\it x$=0.034 of Co in accordance with resistance and magnetization measurements.

Thermoelectric power data as function of temperature for Ba(Fe$_{1-x}$Rh$_x$)$_2$As$_2$, ($0\le x \le 0.171$), are shown in Fig. 5. The onset of superconductivity is observed for x=0.026 of Rh substitution level. Superconductivity is not observable beyond $\it x$=0.13 which is consistent with resistance and magnetization measurements. The TEP data for Rh substitution are very similar to TEP of Co substitution. Although the absolute value of TEP overall is slightly smaller for Rh substituted samples compared to those of Co substituted samples. The thermoelectric power is negative for all concentration range meaning negative charge carriers are dominant in the thermoelectric transport.

\section{Discussion}

\begin{figure}[t]
\centering
\includegraphics[width=1\linewidth]{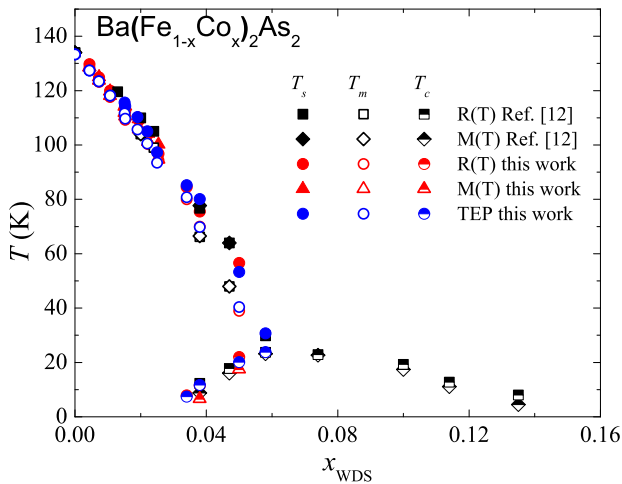}
\caption{\footnotesize (Color online) $\it T- x$ phase diagram of  Ba(Fe$_{1-x}$Co$_x$)$_2$As$_2$. Data for $T_c$, $T_s$ and $T_m$ obtained from resistivity (R) and susceptibility (M)  are adopted from Ref. \cite{Ni}. }
\label{fig:Co-PD}
\end{figure}

\begin{figure}[t]
\centering
\includegraphics[width=1\linewidth]{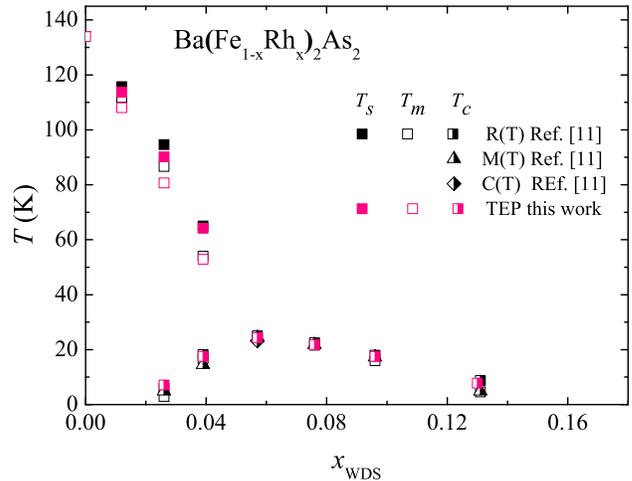}
\caption{\footnotesize(Color online) $\it T- x$ phase diagram of  Ba(Fe$_{1-x}$Rh$_x$)$_2$As$_2$. Data for $T_c$, $T_s$ and $T_m$ obtained from resistivity (R), susceptibility (M) and specific heat (C) are adopted from Ref. \cite{Ni1}.}
\label{fig:PD Rh}
\end{figure}

\begin{figure}[tbh]
\centering
\includegraphics[width=0.9\linewidth]{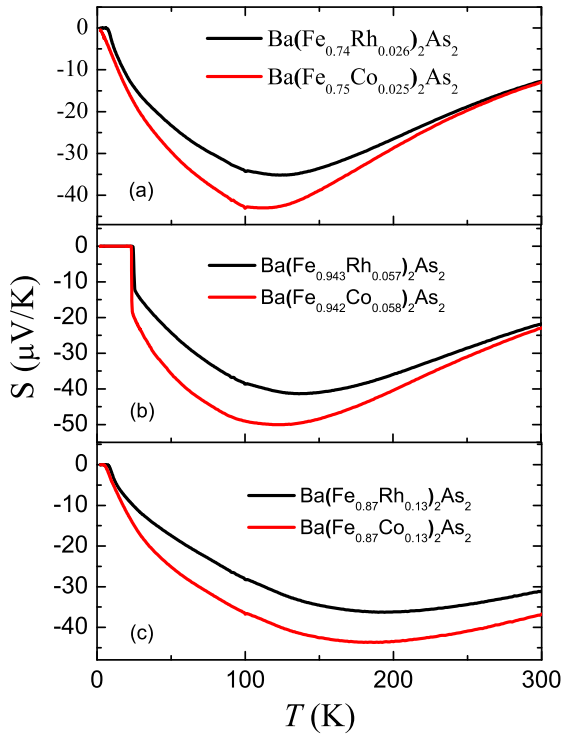}
\caption{\footnotesize (Color online) (a), (b) and (c) In-plane temperature-dependent TEP of selected Ba(Fe$_{1-x}$Rh$_x$)$_2$As$_2$ and Ba(Fe$_{1-x}$Co$_x$)$_2$As$_2$ single crystals with similar $\it x$ values. (b) and (c) TEP data for Ba(Fe$_{1-x}$Co$_x$)$_2$As$_2$ are adopted from Ref. \cite{Eundeok1,Hodovanets} respectively.}
\label{fig:comparison Rh and Co}
\end{figure}

\begin{figure}[tbh]
\centering
\includegraphics[width=1\linewidth]{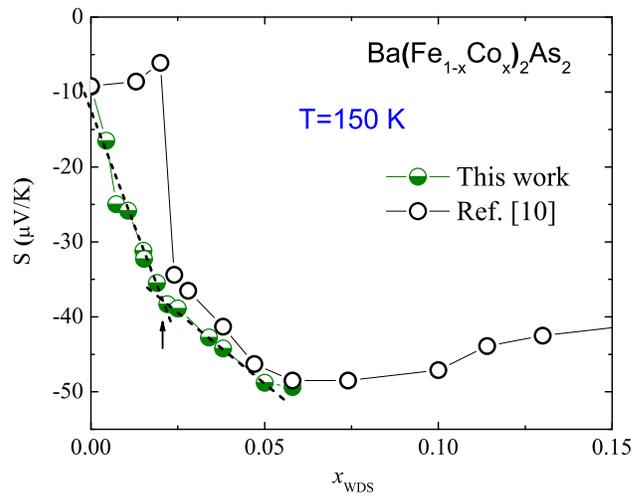}
\caption{\footnotesize (Color online) TEP, $\it S$, as a function of Co concentration, $\it x$, at 150 K. Open symbols data are adopted from Ref. \cite{Eundeok1}. An arrow marks the regions of anomalous $S(x)|_{T=150K}(x)$ behavior. The dashed lines are guides for the eyes. }
\label{fig:TEP analysis}
\end{figure}

\begin{figure}[tbh]
\centering
\includegraphics[width=1\linewidth]{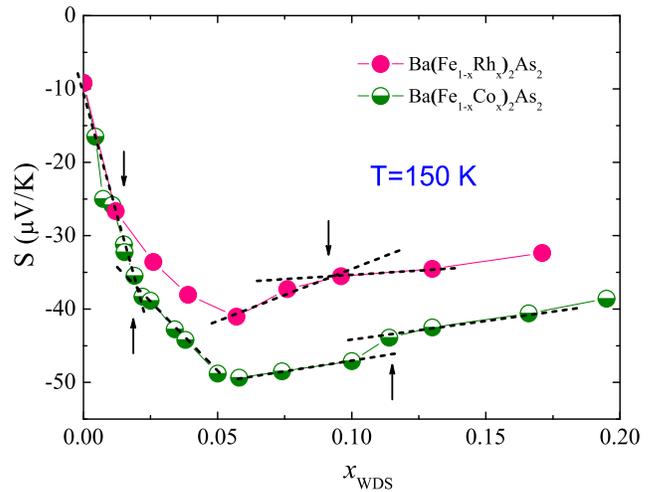}
\caption{\footnotesize (Color online) $\it S$ as a function of Co and Rh concentrations at 150 K. Arrows mark the regions of anomalous $S(x)|_{T=150K}(x)$ behavior. The dashed lines are guides for the eyes. The TEP data for Ba(Fe$_{1-x}$Co$_x$)$_2$As$_2$, ($x > 0.58$) are adopted from Ref.\cite{Eundeok1,Hodovanets}.}
\label{fig:S(x) Co and Rh}
\end{figure}

The structural, $T_s$, magnetic, $T_m$, and superconducting, $T_c$, transition temperatures obtained from temperature dependent TEP data for Ba(Fe$_{1-x}$Co$_x$)$_2$As$_2$, ($0\le x \le 0.05$) and Ba(Fe$_{1-x}$Rh$_x$)$_2$As$_2$, ($0\le x \le 0.171$), Figs. 6 and 7 respectively, are consistent with the phase diagrams constructed from the data obtained by resistivity, magnetization and specific heat measurements \cite{Ni,Ni1}. For $T_c$, an offset in S(T) was used. The criteria for extracting $T_s$ and $T_m$ were adopted from Ref. \cite{Eundeok1}. 

Given that the temperature dependent resistance and magnetization data for Ba(Fe$_{1-x}$Rh$_x$)$_2$As$_2$ are strikingly similar to those of Ba(Fe$_{1-x}$Co$_x$)$_2$As$_2$ resulting in the phase diagrams that are almost identical \cite{Ni}, it is interesting to compare the effects of Rh substitution to those of Co on the temperature-dependent TEP. Figure 8 shows a comparison of the temperature-dependent TEP for selected Ba(Fe$_{1-x}$Rh$_x$)$_2$As$_2$ and Ba(Fe$_{1-x}$Co$_x$)$_2$As$_2$ single crystals. It can be seen, that the temperature dependence of the TEP data for Rh substitution levels is very similar to those for Co substitution levels although the absolute value of thermoelectric power overall is slightly smaller for Rh substituted samples compared to those of Co substituted samples. This difference might be caused by subtle differences in how Co and Rh change the details of the density of states near the Fermi level, and/or a differences in the scattering potential, and/or changes in phonon spectrum (Rh is a heavier than Co atom with larger ionic radius and Rh substitution also affects the lattice parameters in different way compared to Co substitution \cite{Ni1}). Taking into account almost identical phase diagrams of Co and Rh substituted BaFe$_2$As$_2$ and isoelectronic nature of substitution, the difference in the absolute value of the TEP between Rh and Co substituted samples is more likely due to the differences in the scattering potential.

TEP data as a function of Co concentration at 150 K are presented in Fig. 9. The data set from Ref. \cite {Eundeok1}, open symbols, shows a significant jump between $\it x$=0.02 and $\it x$=0.024. On the other hand, the data set from this work's tighter spaced Co substitution levels, shows rather gradual evolution of S(x) with Co substitution with what appears to be a change of slope at around $\it x$=0.02 of Co which maybe associated with the Lifshitz transition observed by ARPES.

If TEP data at fixed temperatures as a function of concentrations are plotted \cite{Varlamov, Blanter}, then possible changes in the Fermi surface topology can sometimes be seen more clearly.  According to the theory, depending on type of the electronic topological transition: void formation or neck disruption \cite{Varlamov, Blanter}, the features in the S versus $\it x$ at constant temperatures will be different. At helium temperatures the  Lifshitz transition might manifest itself as a peak in $|S|$ versus $\it x$ plot. The peak becomes broader as the temperature or amount of impurities is increased. The parameters that affect the width of the peak and the finite values of the thermopower at the peak itself are currently unknown.

In the case of Ba(Fe$_{1-x}$Co$_x$)$_2$As$_2$ and Ba(Fe$_{1-x}$Rh$_x$)$_2$As$_2$ it is not suitable to plot the data at the helium temperature because SC exists for temperatures up to 24 K for some of the Co and Rh substitution levels. In addition, for any temperature cuts below 134 K, below $T_s$/$T_m$ for the parent compound, the S(x) data will have a feature at the $T_s$/$T_m$ line crossing; and below the structural and magnetic transitions, there is magnetic contribution to the thermoelectric power which would be difficult to disentangle from electronic contribution and which (as opposed to phonon contribution) could have significant concentration dependence. Thus, in Fig. 9, 150 K temperature was chosen so that any features associated with the structural and magnetic transitions are not contributing to the data plotted. 

Ba(Fe$_{1-x}$Co$_x$)$_2$As$_2$ is a complex 5-band system and if one or a few bands undergo an electronic topological transition it is unclear what the experimental signature of the Lifshitz transitions would look like. Even if we assume that a Lifshitz transition in a single band would manifest itself as a peak in S versus $\it x$ at very low temperatures, the combination of multiple bands and intermediate temperatures can be expected to mare the signature of a Lifshits transition \cite{Egorov}. This being said, the data from this work shown in Fig. 9 manifest a clear break in slope near $\it x$=0.02, the value identified by earlier ARPES measurements \cite{Liu,Liu1}. 

In Figure 10 we plotted S($\it x$) at 150 K for both Co and Rh substituted BaFe$_2$As$_2$ single crystals for $0\le x \le 0.171$. The data for Co substituted samples are the same as in Fig. 9 with the data for Co concentrations $x>0.058$ taken from Refs. \cite{Eundeok1,Hodovanets}. At a first glance, there appears to be a change in the slope at $x\sim 0.015$ of Rh in S($\it x$) at 150 K. Clearly, more measurements need to be done on a tighter spaced Rh substitution levels in order to show that indeed the Lifshitz transition occurs near this concentration of Rh. The anomaly in the $\it S(x)$ at $x\sim 0.1$ of Co, near the end of the superconducting dome, is associated with the Lifshitz transition observed by ARPES \cite{Liu1}. An anomaly in $\it S(x)$ for Rh substituted sample is also observable just below this concentration and is marked with an arrow in Fig 10. Another interesting features to point out in S(x) are the lowest points in S(x) for both Co and Rh at around $\it x$=0.06. For this concentration the $T_s/T_m$ "lines" coinciding with $T_c$ "line", near optimal substitution and the highest $T_c$. This feature might be associated with significant changes of the electronic structure or other correlations as well as possible quantum critical fluctuations \cite{Arsenijevic}.

\section{Conclusion}
Resistance, magnetization, and thermoelectric power measurements on tightly spaced Ba(Fe$_{1-x}$Co$_x$)$_2$As$_2$, ($0\le x \le 0.05$), and thermoelectric power measurements on Ba(Fe$_{1-x}$Rh$_x$)$_2$As$_2$, ($0\le x \le 0.171$), single crystals are reported. TEP data for Ba(Fe$_{1-x}$Co$_x$)$_2$As$_2$ evolve gradually with Co substitution, instead of a large jump in the TEP data from $\it x$=0.02 to $\it x$=0.024 reported earlier \cite{Eundeok1}. The Lifshitz transition, observed by ARPES, appears to manifests itself as change in slope at $x\sim 0.02$ of Co in S($\it x$) at 150 K. The temperature dependent TEP data of Ba(Fe$_{1-x}$Rh$_x$)$_2$As$_2$ show very similar behavior to that of the Ba(Fe$_{1-x}$Co$_x$)$_2$As$_2$. Change of slope in $S|_{T=150K}(x)$ data for Ba(Fe$_{1-x}$Rh$_x$)$_2$As$_2$, that might indicate at the Lifshitz transition, occurs at concentrations similar to that in the Ba(Fe$_{1-x}$Co$_x$)$_2$As$_2$ series. An anomaly in $S|_{T=150K}(x)$ data at $x\sim 0.1$ for Ba(Fe$_{1-x}$Rh$_x$)$_2$As$_2$ might be associated with Lifshitz transition similar to Co substitution. The previously outlined $\it T - x$ phase diagrams for both systems are confirmed.

\section{Acknowledgments}
The authors would like to thank: W.E. Straszheim for the elemental analysis of the single crystals; M. A. Tanatar for soldering Sn to the ends of the sample; A. Jesche, X. Lin, and V. Taufour as well as the five Kims: H, M, R, S and -chi for valuable discussions. German D. Samolyuk is thanked as well for suggestions for refinement of measurement procedures. This work was done at Ames Laboratory, US DOE, under contract $\#$DE-AC02-07CH111358. This work was supported by the US Department of Energy, Office of Basic Energy Science, Division of Materials Sciences and Engineering. S.L.B. and P.C.C were supported in part by the State of Iowa through the Iowa State University.

\end{document}